\documentclass[11pt]{article}
\usepackage{moriond,epsfig}
\usepackage{amsmath}
\usepackage{amssymb}
\usepackage{amsfonts}
\usepackage{subfig}
\usepackage{hyperref}

\def\Journal#1#2#3#4{{#1} {\bf #2}, #3 (#4)}

\def\PRL{\em Phys. Rev. Lett.}
\def\PRD{{\em Phys. Rev.} D}

\def\be{\begin{equation}}
\def\ee{\end{equation}}
\def\bea{\begin{eqnarray}}
\def\eea{\end{eqnarray}}

\begin{document}
\vspace*{4cm}
\title{PERTURBATIVE, POST-NEWTONIAN, AND GENERAL RELATIVISTIC DYNAMICS OF BLACK HOLE BINARIES}

\author{A. LE TIEC}

\address{Maryland Center for Fundamental Physics \& Joint Space-Science Institute, \\ Department of Physics, University of Maryland, College Park, MD 20742, USA}

\maketitle\abstracts{The orbital motion of inspiralling and coalescing black hole binaries can be investigated using a variety of approximation schemes and numerical methods within general relativity: post-Newtonian expansions, black hole perturbation theory, numerical relativity, and the effective-one-body formalism. We review two recent comparisons of the predictions from these various techniques. Both comparisons rely on the calculation of a coordinate invariant relation, in the case of non-spinning binary black holes on quasi-circular orbits. All methods are shown to agree very well in their common domain of validity.}

\section{Introduction}

The detection and analysis of the gravitational radiation from black hole binaries by the ground-based LIGO/Virgo observatories, and future space-based antennas, requires very accurate theoretical predictions for use as gravitational wave templates. The orbital motion of such compact binary systems can be analyzed using multiple approximation schemes and numerical methods in general relativity: post-Newtonian (PN) expansions, black hole perturbation theory, the effective-one-body (EOB) formalism, and numerical relativity (NR). It is crucial to compare the predictions from these various techniques for several reasons: such comparisons (i) provide independent consistency checks of the validity of the various calculations, (ii) they help to delineate the respective domains of validity of each method, and (iii) they can inform the development of a universal semi-analytical model of the binary dynamics and gravitational wave emission. In this paper, we shall summarize the main results of two such recent comparisons, both related to the {\em local} orbital dynamics of non-spinning black hole binaries on quasi-circular orbits.

\section{Redshift Observable}\label{Sec2}

Our first comparison is concerned with the relativistic motion of compact binary systems within black hole perturbation theory and the PN approximation. Consider two non-spinning black holes with masses $m_1$ and $m_2$, moving on an exactly circular orbit with angular frequency $\Omega_\varphi$. The dissipative effects associated with the emission of gravitational radiation are neglected, which is formalized by assuming the existence of a helical Killing vector (HKV) field $k^\alpha$. The 4-velocity $u_1^\alpha$ of the ``particle'' $m_1$ is necessarily tangent to the HKV evaluated at that location; hence $u_1^\alpha = U \, k_1^\alpha$. The scalar $U$ is a constant of the motion associated with the helical symmetry. It also measures the gravitational redshift of light rays emitted from $m_1$, and received at large distance, along the helical symmetry axis perpendicular to the orbital plane;~\cite{De.08} we shall henceforth refer to $U$ as the ``redshift observable''. Being coordinate invariant, the relation $U(\Omega_\varphi)$ provides a handy testbed to compare the predictions from the two approximation schemes.

For an extreme mass ratio black hole binary, such that $m_1 \ll m_2$, the redshift observable $U(\Omega_\varphi;m_1,m_2)$ is conveniently expanded in powers of the mass ratio $q \equiv m_1 / m_2$, according to
\be\label{eq:U}
    U = U_\text{Schw} + q \, U_\text{GSF} + \mathcal{O}(q^2) \, .
\ee
All coefficients in the expansion \eqref{eq:U} are functions of the dimensionless coordinate invariant PN parameter $y \equiv (m_2 \Omega_\varphi)^{2/3}$. The result for a test mass in circular orbit around a Schwarzschild black hole of mass $m_2$ is known in closed form as $U_\text{Schw} = \left( 1 - 3 y \right)^{-1/2}$. The invariant relation $U_\mathrm{GSF}(y)$ encoding the first order mass ratio correction has been computed numerically, with high precision.~\cite{De.08,Bl.al.10} This  gravitational self-force (GSF) effect has also been computed analytically up to high PN orders.~\cite{Bl.al.10,Bl.al2.10} The post-Newtonian expansion of $U_\text{GSF}$ is of the form
\begin{equation}\label{eq:U_GSF}
	U_\text{GSF} = \sum_{k \geqslant 0} \alpha_k \, y^{k+1} + \ln{y} \sum_{k \geqslant 4} \beta_k \, y^{k+1} + \cdots \, ,
\end{equation}
where the coefficients $\alpha_k$ and $\beta_k$ are pure numbers, and the dots stand for terms involving powers of logarithms $(\ln{y})^p$, with $p \geqslant 2$, which are expected not to occur before the very high 7PN order.~\cite{Bl.al2.10} The Newtonian, 1PN, 2PN and 3PN polynomial coefficients $\{\alpha_0,\alpha_1,\alpha_2,\alpha_3\}$ were determined analytically,~\cite{De.08,Bl.al.10} as well as the leading-order 4PN and next-to-leading order 5PN logarithmic coefficients $\{\beta_4,\beta_5\}$.~\cite{Bl.al2.10} Their values are reported in the left panel of Table~\ref{tab:coeffs}.

\begin{table}[h!]
	\caption{\label{tab:coeffs} \footnotesize The analytically determined post-Newtonian coefficients $\alpha_k$ and $\beta_k$ (left panel), and the numerically determined values of higher-order PN coefficients, based on a fit to the GSF data (right panel). The uncertainty in the last digit is indicated in parenthesis.}
	\vspace{0.15cm}
	\centering
	\begin{tabular}{cc}
		\hline\hline
		Coeff. & Value \\
		\hline
		$\alpha_0$ & $-1$ \\
		$\alpha_1$ & $-2$ \\                 
		$\alpha_2$ & $-5$ \\                 
		$\alpha_3$ & $-\frac{121}{3}+\frac{41}{32}\pi^2$ \\
		$\beta_4$ & $-\frac{64}{5}$ \vphantom{\rule{0pt}{12pt}} \\
		$\beta_5$ & $+\frac{956}{105}$ \vphantom{\rule{0pt}{12pt}} \\
		\hline\hline
	\end{tabular}
	\hspace{1cm}
	\begin{tabular}{cl}
		\hline\hline
		Coeff. & \hspace{0.5cm} Value \\
		\hline
		$\alpha_4$ & $-114.34747(5)$ \\
		$\alpha_5$ & $-245.53(1)$ \\                 
		$\alpha_6$ & $-695(2)$ \\                 
		$\beta_6$ & $+339.3(5)$ \\
		\hline\hline
	\end{tabular}
\end{table}

Making use of the known results for the coefficients $\{\alpha_0,\alpha_1,\alpha_2,\beta_4,\beta_5\}$, a fit to the GSF data for $U_\text{GSF}(\Omega_\varphi)$ gave the numerical estimate $\alpha_3^\text{fit} = - 27.6879035(4)$ for the 3PN coefficient,~\cite{Bl.al.11} to be compared with the exact value $\alpha_3 = - 27.6879026\cdots$.~\cite{Bl.al.10} The results are in agreement with {\em nine} significant digits, at the $2 \sigma$ level. This provides a strong and independent test of the validity of both calculations, which rely on very different regularization schemes to subtract the divergent self-fields of point particles (mode-sum regularization in the self-force, and dimensional regularization in PN theory). ‏By fitting the accurate GSF data to a PN model of the form \eqref{eq:U_GSF}, now taking into account all known PN coefficients, including the exact value of the 3PN coefficient $\alpha_3$, the values of previously unknown PN coefficients $\alpha_k$ and $\beta_k$ were measured, up to the very high 6PN order.~\cite{Bl.al2.10} These are reported in the right panel of Table~\ref{tab:coeffs}. Notice in particular how the 4PN and 5PN coefficients $\alpha_4$ and $\alpha_5$ could be determined with high precision.

\begin{figure}[t!]
	\begin{center}
		\includegraphics[width=9cm]{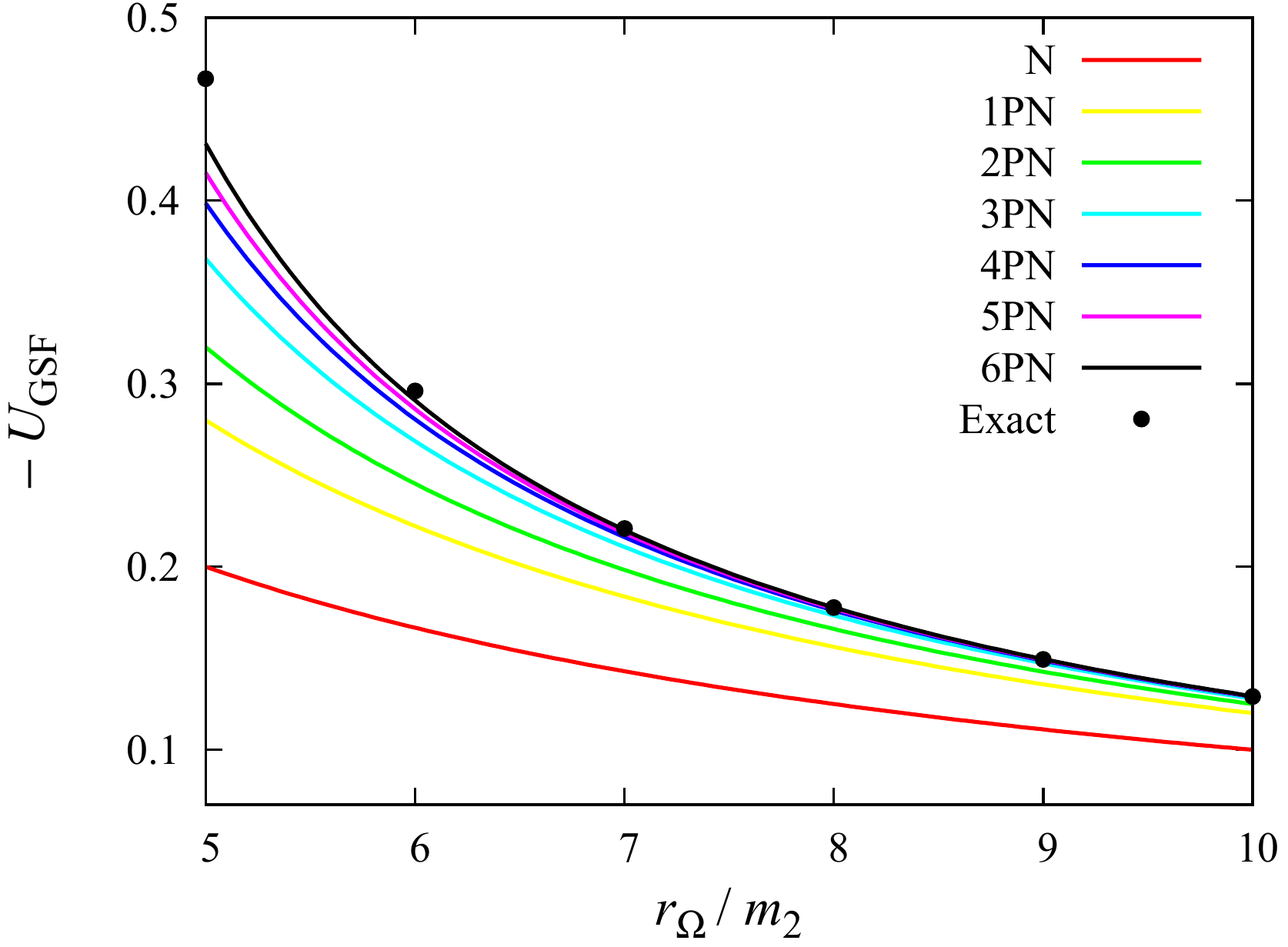}
	\end{center}
	\caption{\label{fig:U} \footnotesize The gravitational self-force contribution $U_\text{GSF}$ to the redshift observable $U$, as a function of $r_\Omega \equiv m_2 / y$, a coordinate invariant measure of the orbital separation. Notice that $r_\Omega = 6 m_2$ corresponds to the very relativistic innermost stable circular orbit (ISCO) of a test-mass orbiting a Schwarzschild black hole of mass $m_2$.}
\end{figure}

Figure~\ref{fig:U} shows the exact results for $U_\text{GSF}(\Omega_\varphi)$, as computed within the self-force, as well as the successive truncated PN series up to 6PN order, based on the analytically and numerically determined PN coefficients summarized in Table~\ref{tab:coeffs}. This comparison illustrates the complementarity of the two approximation schemes: previous knowledge of analytically determined ``low'' order PN coefficients allows to extract from the accurate GSF data information about higher order PN effects, which otherwise would likely remain inaccessible to standard PN calculations.

\section{Periastron Advance}

As long as the radiation-reaction time scale is much longer than the typical orbital time scale, the motion of two non-spinning black holes on a generic eccentric orbit depends on two independent frequencies: the radial frequency (or mean motion) $\Omega_r = 2 \pi / P$, where $P$ is the radial period, i.e. the time interval between two successive periastron passages, and the periastron precession frequency $\Delta \Phi / P$, where $\Delta \Phi / (2 \pi) \equiv K-1$ is the fractional advance of the periastron per radial period. In the zero eccentricity limit, the relation between the circular orbit frequency $\Omega_\varphi$ and $K = \Omega_\varphi / \Omega_r$ is coordinate invariant; it can thus be used as a convenient reference for comparison.

The invariant relation $K(\Omega_\varphi)$ has been computed at the 3PN accuracy in PN theory,~\cite{Da.al.00} at first order in perturbation theory,~\cite{Ba.al.10} and in the EOB formalism.~\cite{Da.10} This genuine general relativistic effect has also recently been measured for the first time in fully non-linear NR simulations.~\cite{Mr.al.10} Le~Tiec~{\em et al}.~\cite{Le.al.11} considerably improved upon the accuracy of this initial measurement. Making use of new and longer simulations of the late stage of the inspiral of non-spinning black hole binaries with mass ratios $q=1,2/3,1/3,1/5,1/6$, and $1/8$,~\cite{Mr.al.11} they measured $K$ with a relative uncertainty $\sim 0.1-1\%$. This accuracy made possible an extensive comparison which, for the first time, (i) encompassed all the analytical and numerical methods currently available, and (ii) focused on the orbital dynamics of the binary, rather than the asymptotic waveform.

\begin{figure}[t!]
	\begin{center}
		\includegraphics[width=15.5cm]{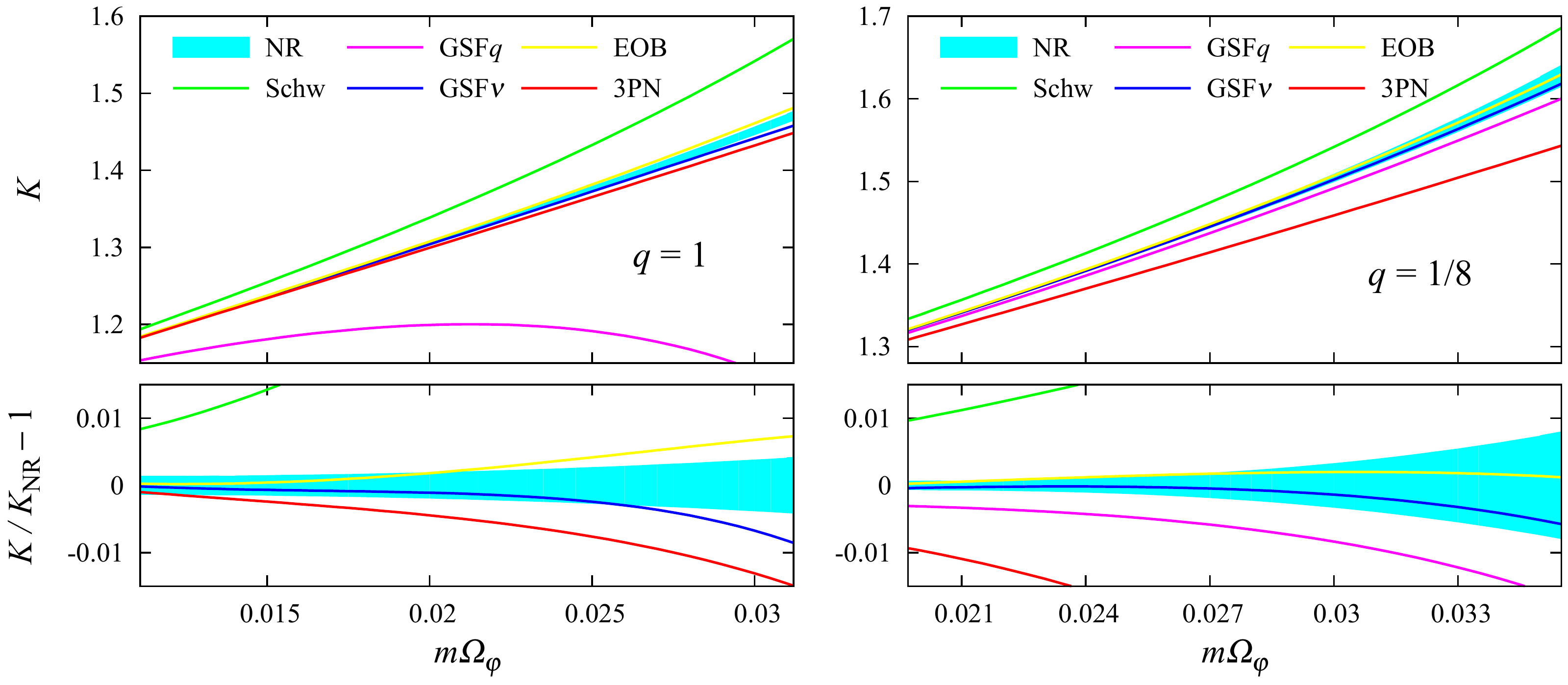}
		\caption{\label{fig:K} \footnotesize The periastron advance $K = 1 + \Delta \Phi / (2 \pi)$, as a function of the circular orbit frequency $\Omega_\varphi$, for black hole binaries with mass ratios $1:1$ (left panel) and $1:8$ (right panel). Notice that in the later case, $\Delta \Phi$ reaches half an orbit per radial period for $m \Omega_\varphi \sim 0.03$, corresponding to an orbital separation $r \sim 10 m$.}
	\end{center}
\end{figure}

Figure~\ref{fig:K} shows the invariant relation $K(\Omega_\varphi)$ for binary black holes with mass ratios $q=1$ (left panel), and $q=1/8$ (right panel), as computed in NR (in cyan), PN theory (red), and the EOB formalism (yellow). For comparable masses (e.g. $q=1$ or $2/3$), the 3PN prediction is in good agreement with the exact result from NR (to better than $1\%$). However, as expected, it performs less well when $q \to 0$.~\cite{Bl.02} The EOB (3PN) prediction, on the other hand, is in very good agreement with the NR data over the entire range of frequencies and mass-ratios considered.

Also shown in Fig.~\ref{fig:K} are the predictions for a test mass in circular orbit around a Schwarzschild black hole (green), and the inclusion of the GSF (magenta and blue). While perturbative self-force calculations are commonly formulated as expansions in powers of the usual mass ratio $q$ [see e.g. Eq.~\eqref{eq:U}], PN expansions naturally involve the {\em symmetric} mass ratio $\nu \equiv m_1 m_2 / m^2$, where $m = m_1 + m_2$ is the total mass of the binary. Since at first order $q = \nu + \mathcal{O}(\nu^2)$, the GSF result for the periastron advance may as well be written in the ``resummed'' form
\be\label{eq:K}
	K = K_\text{Schw} + \nu \, K_\text{GSF} + \mathcal{O}(\nu^2) \, ,
\ee
where all coefficients are functions of the dimensionless invariant PN parameter $x \equiv (m \Omega_\varphi)^{2/3}$. The GSF correction $K_\text{GSF}$ to the test-particle result $K_\text{Schw} = \left( 1 - 6 x \right)^{-1/2}$ has recently been computed numerically.~\cite{Ba.al.10} Although the GSF$q$ prediction [obtained by replacing $\nu \to q$ in Eq.~\eqref{eq:K}] agrees with the exact result within a relative difference of magnitude $\sim q^2$, as expected, the GSF$\nu$ prediction \eqref{eq:K} agrees remarkably well with the NR data for {\em all} mass ratios. This surprising result suggests that GSF calculations may very well find application in a broader range of physical problems than originally envisaged, including the modelling of intermediate mass ratio inspirals, a plausible source of gravitational waves for Advanced LIGO/Virgo.

\section*{Acknowledgments}

The results summarized in this paper were obtained in collaboration with L.~Barack, L.~Blanchet, A.~Buonanno, S.~Detweiler, A.H.~Mrou{\'e}, H.P.~Pfeiffer, N.~Sago, A.~Taracchini, and B.F.~Whiting. The author acknowledges support from NSF Grant PHY-0903631, and from the Maryland Center for Fundamental Physics. He is grateful to the organizing committee of the {\em 46th Rencontres de Moriond \& GPhyS Colloquium} for their kind invitation, and for providing financial support.


\section*{References}

\begin{thebibliography}{99}
\bibitem{De.08} S. Detweiler, \Journal{\PRD}{77}{124026}{2008}.
\bibitem{Bl.al.10} L. Blanchet, S. Detweiler, A. Le Tiec, and B.F. Whiting, \Journal{\PRD}{81}{064004}{2010}.
\bibitem{Bl.al2.10} L. Blanchet, S. Detweiler, A. Le Tiec, and B.F. Whiting, \Journal{\PRD}{81}{084033}{2010}.
\bibitem{Bl.al.11} L. Blanchet, S. Detweiler, A. Le Tiec, and B.F. Whiting, in {\em Mass and Motion in General Relativity}, eds. L. Blanchet, A. Spallicci, B. Whiting (Springer, 2011).
\bibitem{Da.al.00} T. Damour, P. Jaranowski, and G. Sch{\"a}fer, \Journal{\PRD}{62}{044024}{2000}.
\bibitem{Ba.al.10} L. Barack, T. Damour, and N. Sago, \Journal{\PRD}{82}{084036}{2010}.
\bibitem{Da.10} T. Damour, \Journal{\PRD}{81}{024017}{2010}.
\bibitem{Mr.al.10} A.H. Mrou{\'e}, H.P. Pfeiffer, L.E. Kidder, S.A. Teukolsky, \Journal{\PRD}{82}{124016}{2010}.
\bibitem{Le.al.11} A. Le Tiec {\em et al.}, \Journal{\PRL}{107}{141101}{2011}.
\bibitem{Mr.al.11} A.H. Mrou{\'e} {\em et al.}, in preparation.
\bibitem{Bl.02} L. Blanchet, \Journal{\PRD}{65}{124009}{2002}.
\end{thebibliography}
\end{document}